\documentclass[12pt,preprint]{aastex}
\usepackage{amsmath}
\usepackage[retainorgcmds]{IEEEtrantools}
\usepackage{comment}
\usepackage[font=footnotesize]{caption}
\usepackage{subfigure}

\begin{document}

\title{The Close Binary Fraction of Dwarf M Stars}
\author{Benjamin M. Clark}
\affil{Penn Manor High School, 100 East Cottage Avenue, Millersville, PA, 17551}
\affil{California Institute of Technology, MSC 235, Pasadena, CA, 91126}

\author{Cullen H. Blake}
\affil{Princeton University, Department of Astrophysical Sciences, Peyton Hall,
Ivy Lane, Princeton, NJ 08544}

\author{Gillian R. Knapp}
\affil{Princeton University, Department of Astrophysical Sciences, Peyton Hall,
Ivy Lane, Princeton, NJ 08544}

\begin{abstract}

We describe a search for close spectroscopic dwarf M star binaries
using data from the Sloan Digital Sky Survey (SDSS)
to address the question of the rate of occurrence of
multiplicity in M dwarfs. We use a template fitting technique 
to measure radial velocities from 
145,888 individual spectra obtained for a magnitude-limited sample of 39,543 
M dwarfs. Typically,
the three or four spectra observed for each star are separated 
in time by less than four hours, but for $\sim$ 17\% of the stars, the
individual observations span more than two days. 
In these cases we are sensitive to large amplitude radial velocity variations on 
time scales comparable to the separation
between the observations. We use a control sample of objects having 
observations taken within a four hour period to make an empirical
estimate of the underlying radial velocity error distribution and simulate our 
detection efficiency for a wide range of binary star systems. 
We find the frequency of binaries among the 
dwarf M stars with $a<0.4\ \text{AU}$ 
to be $3-4\%$. Comparison with other samples of
binary stars demonstrates that the close 
binary fraction, like the total binary fraction, is an increasing 
function of primary mass.

\end{abstract}

\section{Introduction}

The study of the orbital motions of stars in multiple systems has been 
an active area of research for more than a century (see \citealt{1986asoh.book.....H}).
The fraction of stars which lie in multiple systems (or conversely the 
fraction that are single), the dependence of these fractions on the mass
of the primary, and the separation and mass ratio distributions 
of systems of multiple stars, are all observable quantities which
can be measured through spectroscopic and imaging surveys and 
provide fundamental information in two important areas of stellar
astrophysics. First, in certain circumstances observations of 
binary star systems directly
measure the masses, radii, and luminosities of stars
and provide the experimental bedrock for much of 
our theoretical understanding of the structure and evolution of stars. 
Second, surveys to determine 
the overall statistical properties of multiple systems within a 
population of stars help constrain their formation history;
a detailed understanding of the fraction
of binary stars across the HR diagram is 
an important part of a comprehensive picture of the process of star 
formation. To date, only a few measurements of the binary fraction 
and distribution for low-mass stars have been made - these stars are 
of low luminosity and the acquisition of the necessary large set
of measurements is more difficult than for stars of higher mass and
luminosity. They are of particular importance, however, because 
their formation and dynamical evolution within the systems in which they
are born spans the mass range roughly between those of solar-type
stars and planets.

The current understanding of star formation is that stars form in
small-N clusters which are then broken apart by (short-term)
dynamical decay and (longer-term) 
dynamical destruction \citep{2007prpl.conf..133G}.
Due to the
instability of multiple systems, some members of the system are
likely to be ejected on a relatively short time
scale; \citet{1986Ap&SS.124..217A} showed that in the vast majority 
of cases, the least massive star is ejected.
On longer time scales, interactions
with other stars in the
star cluster disrupt binary stars
\citep{2007prpl.conf..133G}. Both of these mechanisms
predict that the multiplicity of stars should decrease to
lower primary mass, but ejection is more important for close systems
while interactions with other stars are more important for wide,
loosely bound, systems.

The properties of multiple star systems with Sun-like primaries have 
received much attention of late. Recent work by 
\citet{raghavan2010}, which updates 
the seminal results of \citet{1991A&A...248..485D}, found that 
Sun-like stars have an overall multiple fraction of $\sim$45\%, with 
a log-normal distribution in separation that peaks around 60 AU. 
A similar analysis of lower mass M stars by \citet{1992ApJ...396..178F} 
found a somewhat lower overall binary frequency of 42$\pm$9\%. For 
objects straddling the stellar--brown dwarf boundary,
\citet{2007ApJ...668..492A} found
that ultracool dwarfs (M6 and later) have a binary frequency of 
20$\pm4$\%, while \citet{2010arXiv1009.4197S} 
report that  $\sim$75\% of O stars
are binaries. Based on these results, \citet{lada2006} 
and \citet{raghavan2010}  point out that
the overall fraction of single stars appears 
to be a decreasing function
of stellar mass. 

\citet{1992ApJ...396..178F} report 
that 1.8$\pm$1.8\% of early-M stars are binaries with 
$0.04\ \text{AU}<a<0.4\ \text{AU}$ based on a sample of 
62 stars, while \citet{2010ApJ...723..684B} report that 
$2.5^{+8.6}_{-1.6}\%$ of late-M and L dwarfs
are binaries with $a<1\ \text{AU}$ based on a sample 
of 43 objects. Extrapolating from the overall separation
distribution functions presented  by \citet{raghavan2010} and
\citet{2010arXiv1009.4197S} provides estimates that
3.7\% and 26\% of G and O stars are binaries with $a<0.4\ \text{AU}$, 
respectively. 
A complete view of stellar multiplicity across a wide range of 
separations, however,
requires the use of a variety of observational techniques. 
Nearby systems with 
wide separations may be directly resolved using  high resolution imaging, 
while systems with small separations are most readily
detected as spectroscopic binaries. In particular, measuring the close binary fraction requires an extensive radial velocity (RV) survey of 
the type described by \citet{raghavan2010}.

Typically, individual multiple systems are identified in radial velocity surveys by 
fitting spectroscopic orbital solutions to velocity curves containing a 
large number of observations. Since close binaries are thought to be 
relatively rare for all but the most massive stars, a rigorous estimate 
of the multiplicity fraction requires a large number of observations 
of a large number of stars over long periods of time
in order to detect an appreciable number of 
systems. This is particularly difficult for dwarf M stars because of their
very low luminosities. An alternative approach is to use a small 
number of observations each for
a very large number of stars to understand the rate 
of occurrence of multiple systems in a statistical sense 
(see \citealt{2005MNRAS.362L..45M}; \citealt{pourbaix05}). 
We take this approach in this paper, taking advantage of 
the enormous numbers of spectroscopic 
observations of M stars gathered during the course of the Sloan 
Digital Sky Survey (SDSS) to measure the close binary fraction of M stars
using observations of 39,543 M dwarfs. While this data set is 
far larger than any used in previous research, it presents
unique challenges. The data produced by the SDSS have significantly 
lower resolution and  signal-to-noise ratio (S/N)
then those typically used for making
stellar velocity measurements. Additionally, SDSS typically only
obtained three or four  spectroscopic exposures of each object; 
in comparison, \citet{1992ApJ...396..178F}
obtained on average 15 observations per star. We demonstrate that an 
overall radial velocity precision of 4 km s$^{-1}$ per observation
can be achieved, more than 
sufficient for the detection of binary systems with short orbital periods, 
even with only few individual measurements. By quantifying the underlying 
statistical properties of the radial velocity measurements extracted from the SDSS spectra,
and simulating the detection efficiency as a function of binary orbital
separation and mass ratio, we make a robust measurement of the 
population of close binary M star systems.

\section{SDSS Spectroscopic Data and Sample Selection}

\subsection{The Sloan Digital Sky Survey}

The Sloan Digital Sky Survey \citep{york00} made an imaging and
spectroscopic survey of the sky using a large-format CCD camera 
\citep{gunn98} mounted on the Sloan Foundation 2.5 m telescope 
\citep{gunn06} at the Apache Point Observatory, New Mexico, to image the sky
in five optical bands - $u$, $g$, $r$, $i$, and $z$ \citep{fukugita96}.
The imaging data are reduced by a set of 
software pipelines which produce a catalog of objects with calibrated
magnitudes and positions (\citealt{lupton01}, \citealt{lupton03}, 
\citealt{hogg01}).
Targets for spectroscopy are selected from this catalog, mapped onto
fiber plug plates \citep{blanton03}, and observed with two dual
fiber-fed spectrographs \citep{uomoto99}  in a series of  several 15-minute
exposures. The spectra are optimally extracted, calibrated, combined
(the {\it combined spectra}), and classified.

\subsection{Selection of the  M Star Sample}

SDSS Data Release 7 (DR7: \citealt{2009ApJS..182..543A}) includes 
spectra of over 1.6 million objects, about 460,000 of which 
are stars.
The parent M-star sample was selected from the spectroscopic data in DR7
by applying a series of cuts, first by magnitude and color ($z \leq 19.5$, $i-z >$ 
0.2 and $r-i > 0.5$). Next, data from plates with diffuse ionized gas
emission and bad spectra (low S/N, missing data etc.) were removed, as were
objects spectroscopically classified as stars earlier than K, as galaxies,
or as quasars. The sample spectra were then examined
by eye.  Stars with obvious blue/white dwarf companions,
stars superimposed on galaxies, and stars
with the spectra of metal-poor (subdwarf M), K, or carbon stars were
removed. 
The resulting sample (Knapp et al. in preparation) contains 51,193
individual M0- L0 stars with spectral types as defined 
by \citet{2005PASP..117..706W}.

\subsection{Multiple SDSS Spectra}

The SDSS usually obtained only one combined spectrum per object, but in a few cases there are two or more
combined spectra, acquired one of two ways: (1) inadvertent observations
of the same object on more than one spectroscopic plate, or (2) duplicate
observations for quality check purposes, either of the same object on two
plates or on the same plate, re-plugged and re-observed. These observations
can be used to search for variability over periods from days to years,
an example being the search by 
\cite{pourbaix05} for spectroscopic binaries. In addition, the
spectrum of each object is a combination of several exposures, and the
spectra obtained from the individual exposures can also be used 
to search for variability, including radial velocity variability,
as is done in the present paper.

The SDSS spectroscopic data are acquired as a series of 
spectroscopic exposures
observed sequentially with an exposure time of 15 minutes each ({\it the 
individual 15 minute spectra}). These spectra
are then averaged to produce the combined spectrum. The data are 
taken this way both to increase the dynamic range of the spectroscopic
observations 
and to allow for the easy removal of 
data artifacts such as ``cosmic rays''. Because of the scientific value
of the individual spectra, software to reduce and calibrate
them was prepared for release in DR7
along with the combined spectra. This data product
is made available to enable studies of rapid spectral variability, since 
it provides, in almost all cases, three spectra observed with the same signal
to noise ratio within a time period of one hour. The science enabled by these
data includes identifying objects with short-period radial 
velocity variations such as compact
degenerate binary pairs \citep{badenes09}, and the variability
of the emission lines in dMe stars \citep{kruse10,hilton10}.

The typical exposure sequence is three fifteen-minute observations
within a time span of about an hour.  However, in marginal weather more
exposures may be required to achieve the required S/N, or the
observational sequence may be interrupted by variable weather
conditions or instrumental problems. In other cases, the required 
S/N may not be achieved 
in a single night, and the individual observations are separated by 
significantly longer times.
This enables searches for spectroscopic variability in the time domain,
analogous to the photometric variability studies that have been carried
out by \citet{blake2008}, \citet{bhatti2010}, and \citet{becker2011}.
The present paper uses the individual spectra to search for 
radial velocity variations in the large sample of M stars defined above. 
We use only the observations from the same plugging of a plate (see
(3) above) and do not consider the repeat observations of 
individual stars or plates (cases 1 and 2 above) - these are
discussed elsewhere.

\section{The Distribution of Radial Velocities}

\subsection{The Final Samples}

We restricted the analysis to stars with $16<i<20.5$.
We broke the sample into sub-samples based on $\Delta t$, the total 
time baseline spanned by the observations of an object: a {\it control sample}
with $0\ \text{hr}\le \Delta t\le 4\ \text{hr}$ and an {\it experimental 
sample} with $2\ \text{d}\le \Delta t\le 30\ \text{d}$. We expect
very few stars in the control sample to undergo significant accelerations
on such short time scales. We use the statistical properties
of the radial velocity measurements of the objects in this sample as an empirical 
estimate of the underlying radial velocity error distribution. Among the set of objects with
$2\ \text{d}\le \Delta t\le 30\ \text{d}$, the time spread between 
the observations is long enough that if the object is a close binary star 
we would observe significant radial velocity variations.

\subsection{Calculation of Radial Velocities}
\label{sec:calcrv}

We use a $\chi^{2}$ minimization technique to estimate radial velocities from the 
SDSS spectra employing the average low-mass star templates 
from  \citet{2007AJ....133..531B} as a zero-velocity reference. 
Using a code written in Interactive Data Language (IDL), we determine 
the radial velocity
that gives the best fit between the spectra and one of the templates 
by minimizing
\begin{equation}
\label{eq1}
\chi^2=\sum_{i }[\frac{f_i-m(\lambda_i)}{\sigma_i}]^2
\end{equation}
where the sum is over all pixels in the spectrum (except for those flagged as bad, see below), $f_i$ is the calibrated 
flux of the $\text{i}^{\text{th}}$ pixel,
$m(\lambda_i)$ is the value of the model at $\lambda_i$ (the wavelength 
of the
$\text{i}^{\text{th}}$ pixel), and $\sigma_i$ is the estimated standard
deviation of $f_i$. We considered spectra 
only from the red arm of the SDSS spectrograph, which spans the region
$\lambda=5800-9200\ \text{\AA}$ at a resolution of 
$\lambda/\Delta\lambda \approx 1800$ and
contains 2048 pixels \citep{2002AJ....123..485S}. Since the 
templates provided by \citet{2007AJ....133..531B} are normalized, 
there are 
two parameters that determine the model $m$ that is the best fit 
for each spectrum: the radial velocity and an overall flux scaling. 
For each spectrum, we found the minimum of the $\chi^2$ curve 
using a iterative process in which
we tested velocities in the range $-1000<RV<1000$ km s$^{-1}$ 
at a resolution of 1 km s$^{-1}$
and then on a finer grid about the velocity found to have the 
smallest $\chi^2$ in the
previous step. This was repeated twice with the final step having a 
resolution of 0.1 m s$^{-1}$.
The templates created by \citet{2007AJ....133..531B} give values 
for the flux at 0.1\AA{} intervals
from 3825\AA{} to 9200\AA. For each test RV, the spectral template is 
interpolated onto a Doppler-shifted wavelength grid using cubic 
spline interpolation \citep{1992nrca.book.....P}. We fit each of 
the 11 templates (one for each spectral class from M0-L0)
to each 15-minute spectrum of each star in our sample in order to 
find the best fitting template, and then select a single template that is
used for all observations of a given target by finding the template 
that results in the lowest total $\chi^{2}$ when fit to all of the 
observations of a given object. We correct for the barycentric motion 
of the Earth by applying velocity corrections produced by the SDSS pipeline.

During the fitting process, we exclude by down weighting spectral 
pixels that might bias the resulting radial velocities. These include any pixels at 
wavelengths greater than $\lambda > 9150 \text{\AA}$, any pixels in 
the H$\alpha$ emission region  $6540  < \lambda < 6585
\text{\AA}$, or pixels with the \textbf{BADSKYCHI} flag set by the 
SDSS pipeline. We exclude the reddest 50\AA{} of spectra since at 
these wavelengths telluric absorption and sky emission become significant
and large radial velocity shifts may extend beyond the edges of the spectral templates. 
The H$\alpha$ emission line can be very strong in active 
M stars and is known to vary significantly even between individual
observations of a given M star (\citealt{2007AJ....133..531B}; 
\citealt{kruse10}), so we chose to exclude pixels within 22\AA{} 
of this feature. 
The \textbf{BADSKYCHI} flag indicates that sky emission lines 
are not being well fit by the SDSS
spectral extraction pipeline, resulting in spectra that are potentially 
contaminated by sky emission. 

\subsection{Identifying Binaries}
\label{sec:id}

Given this sample of more than 145,000 radial velocity measurements, we hope 
to quantify our ability to detect short-timescale radial velocity variability in these data
and estimate the rate of occurrence of short-period binary systems. 
Before doing this, we apply several cuts to the sample. We retain 
only radial velocity measurements from spectra that satisfy 
all of the following criteria: the average S/N of the 
pixels is greater than 10,
the observation is not among the 10\% of observations with the 
largest values of the
$\chi^2$ for the template fit, and
after applying the two previous cuts, the spectra are of objects with at 
least three observations.
We also remove objects from the sample that no longer fall into the 
definitions of our control and experimental samples due to the removal of 
observations by the previous cuts.

\begin{figure}[!ht]
\centering

\subfigure[]
{\includegraphics[scale=.4]{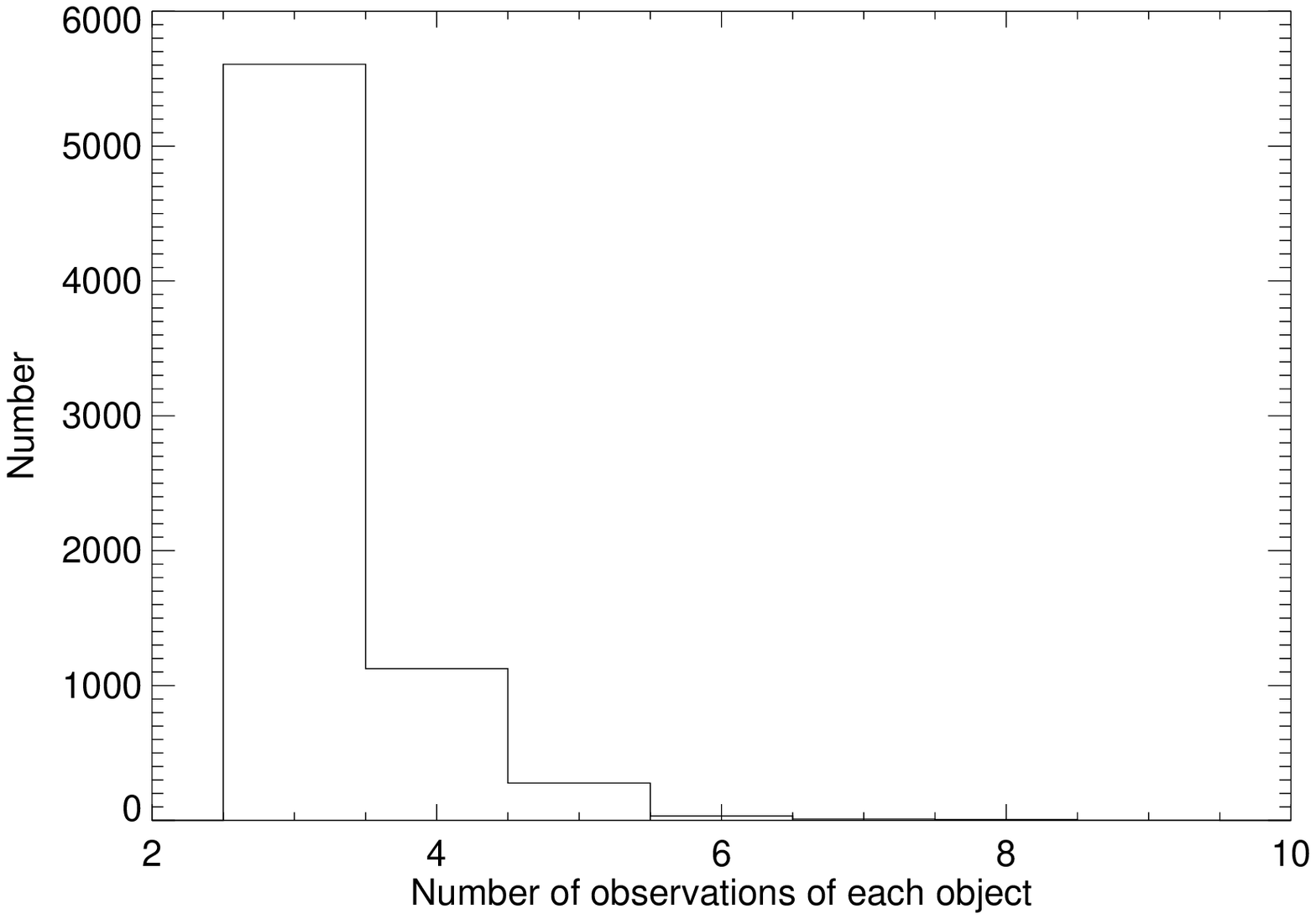}
\label{fig:numspec0_4hr}
}
\subfigure[]
{\includegraphics[scale=.4]{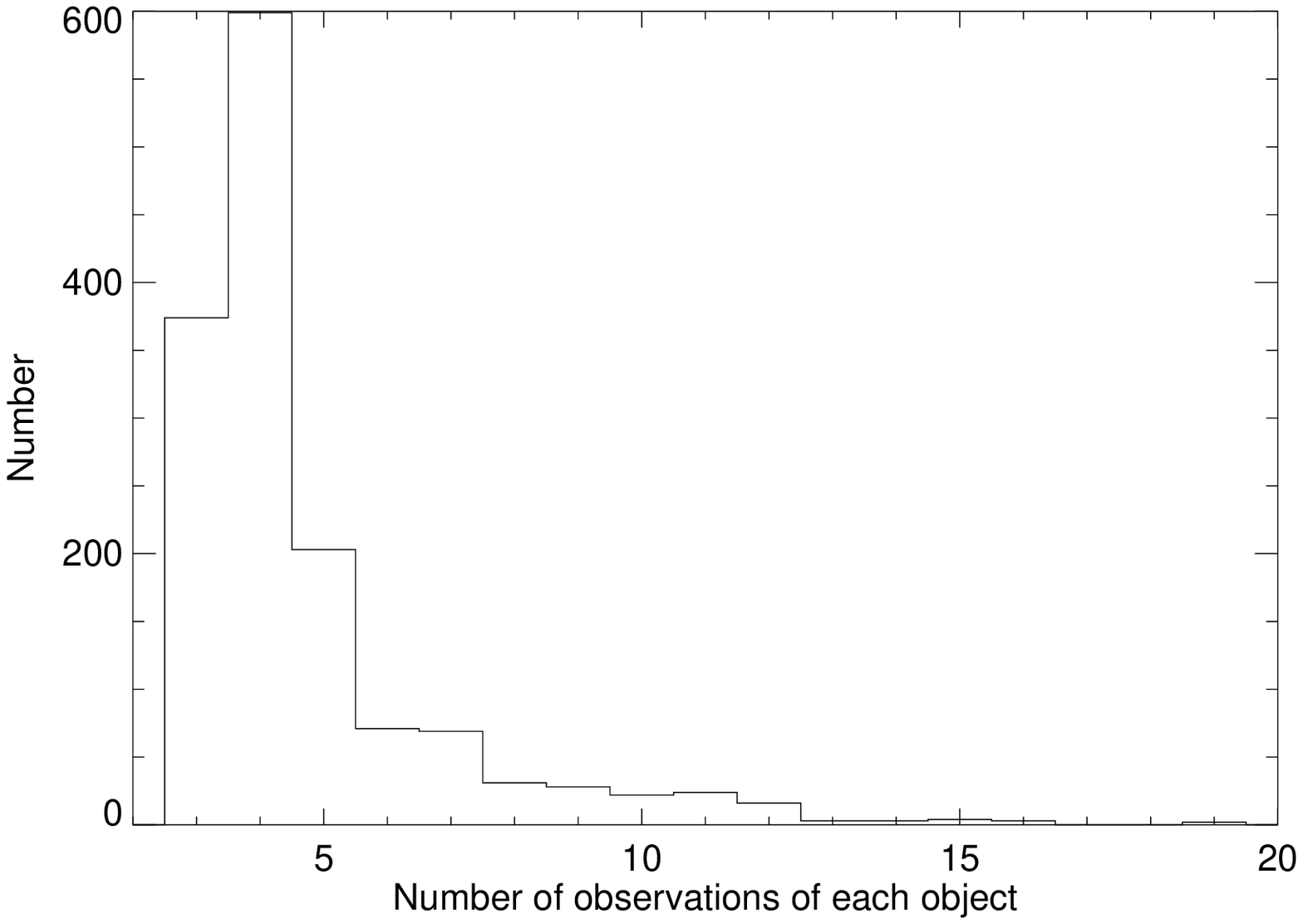}
\label{fig:numspec2_30d}
}

\subfigure[]
{\includegraphics[scale=.4]{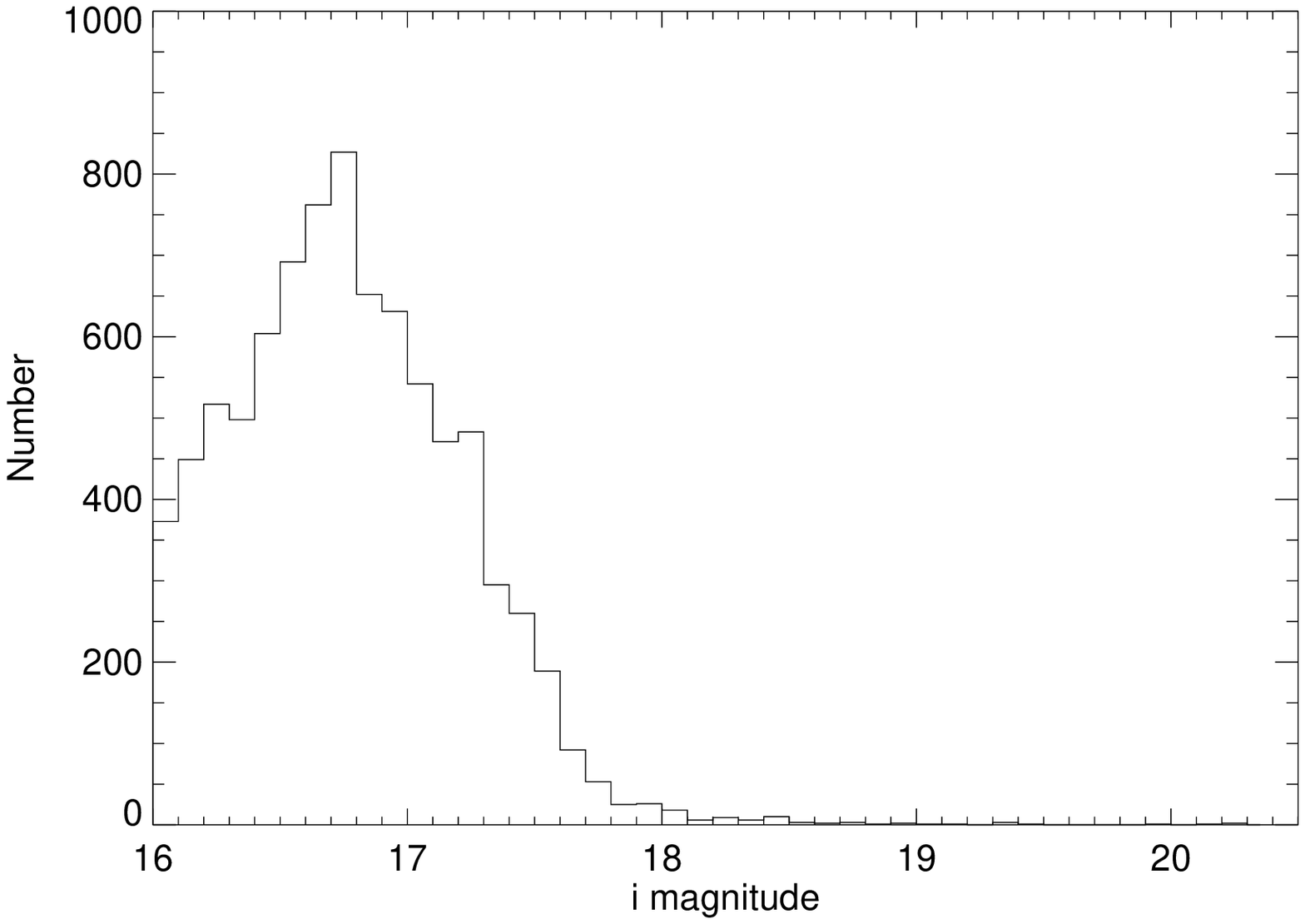}
\label{fig:imag_distibution}
}
\subfigure[]
{\includegraphics[scale=.4]{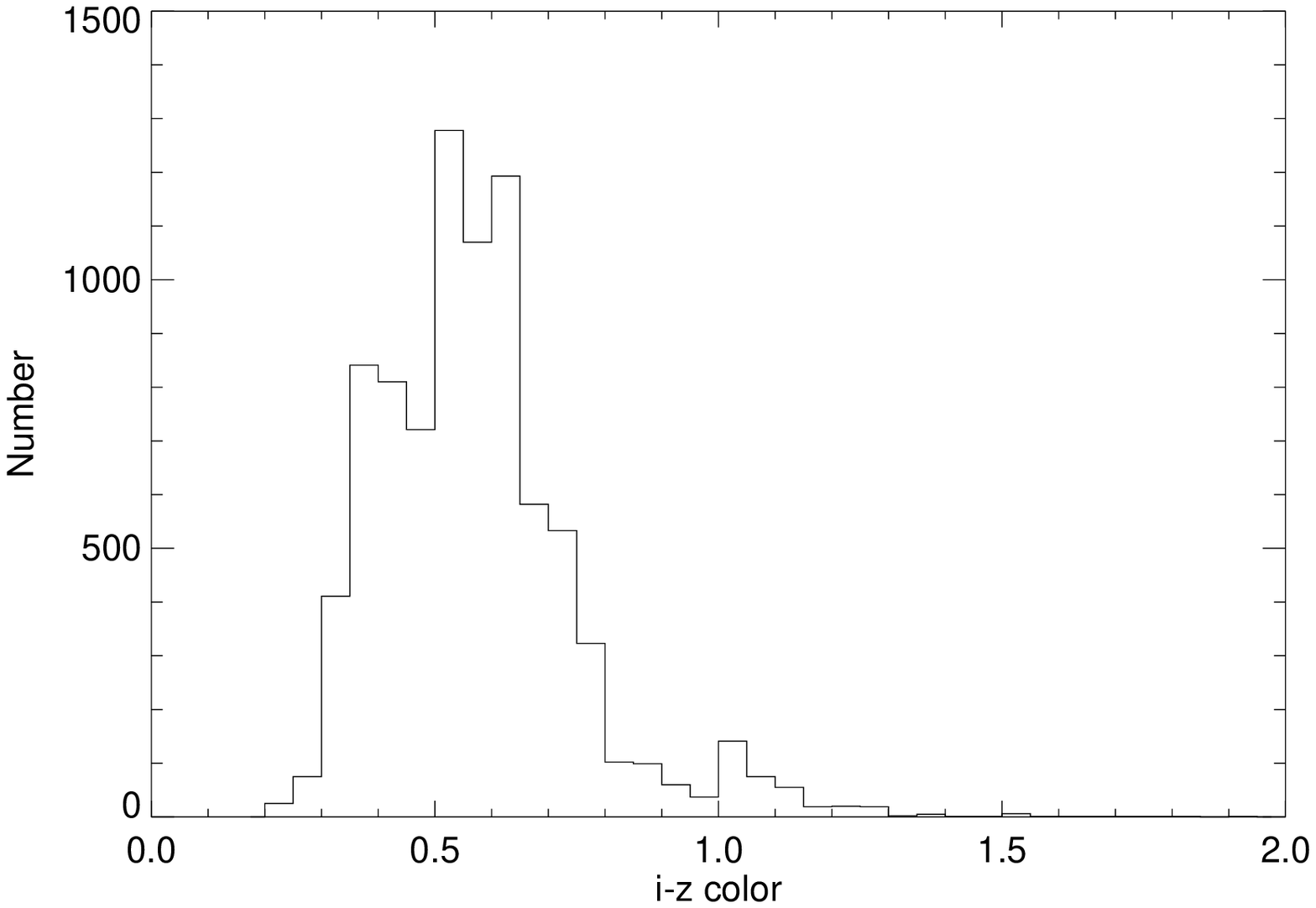}
\label{fig:iz_color_dsitribution}
}
\subfigure[]
{\includegraphics[scale=.4]{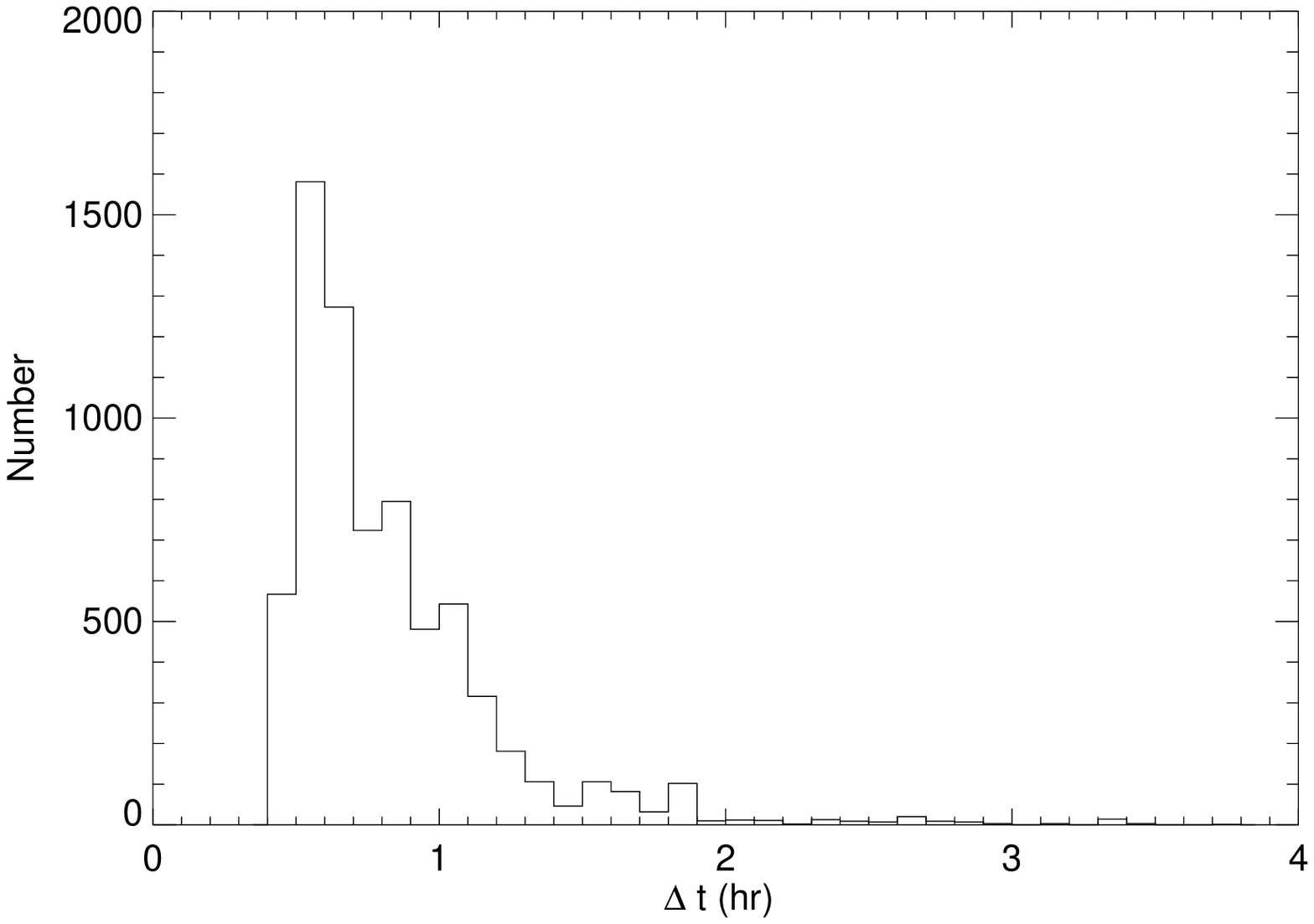}
\label{fig:deltat0_4hr}
}
\subfigure[]
{\includegraphics[scale=.4]{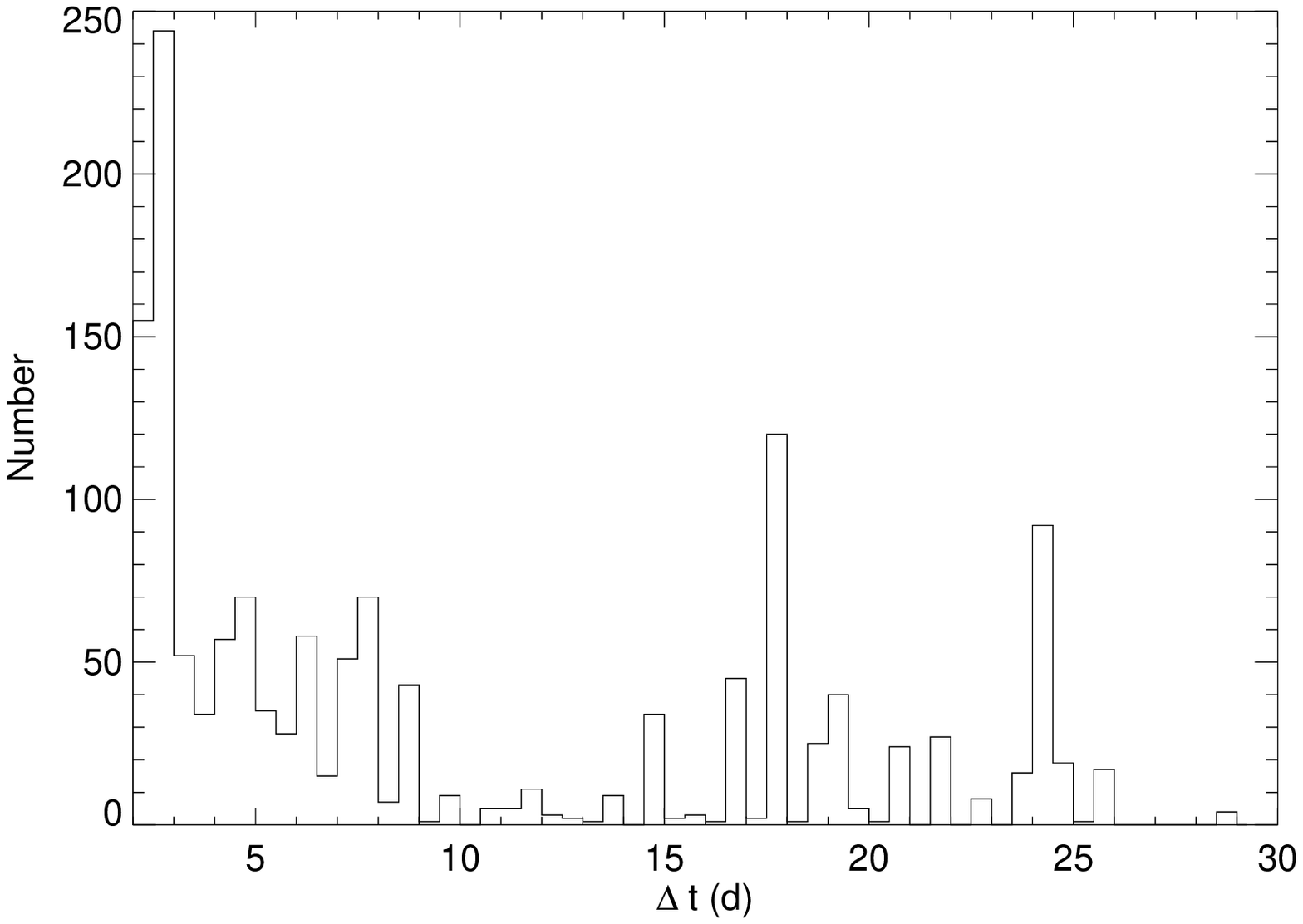}
\label{fig:deltat2_30d}
}
\caption{Subfigures (a)--(f): histograms of the number of observations of each object in the sample
with $0\ \text{hr}\le \Delta t\le 4\ \text{hr}$, the number of observations of each object in the sample with
$2\ \text{d}\le \Delta t\le 30\ \text{d}$, the $i$ magnitude in the combined sample, the $i-z$ color in
the combined sample, $\Delta t$ in the sample with $0\ \text{hr}\le \Delta t\le 4\ \text{hr}$, and
$\Delta t$ in the sample with $2\ \text{d}\le \Delta t\le 30\ \text{d}$, respectively.}
\label{fig:sample_histograms}
\end{figure}

After applying these cuts, 23,031 observations of 7,059 objects remain in 
the control sample and 6,845 observations of 1,452 objects 
in the experimental sample; thus, of the final set of objects analyzed,
17\% have observations with a time span of two days or more. In Figure
\ref{fig:sample_histograms} we show the distributions of the number of
observations of each object and $\Delta t$ 
for both samples, as well as the distributions of $i$-band magnitudes 
and $i-z$ colors. 

For each of the objects in both samples, we define relative velocities, 
$\Delta RV_{i}$,
\begin{equation}
\Delta\text{RV}_i=\text{RV}_i-\frac{\sum_i \text{RV}_i \cdot 
(\text{S/N})_i}{\sum_i (\text{S/N})_i}
\end{equation}
where $\text{RV}_i$ and $(\text{S/N})_i$ represent the RV, after applying
the barycentric correction, and the average signal-to-noise ratio of
all pixels of the
$\text{i}^{\text{th}}$ observation of the object, respectively. 
This is the difference between the radial velocity of the
observation and the weighted average of the radial velocities of all 
observations where the
weight function is $(\text{S/N})_i$.

Given only a small number of observations of an object, it is extremely 
important to understand the underlying radial velocity error distribution if we hope
to reliably detect radial velocity variations. For example, systematic sources of 
error can result in significant non-Gaussian tails to the error distribution
that could produce spurious detections of radial velocity variability. Instead of 
relying on statistical estimates of the radial velocity error for a given measurement,
we use the distribution of $\Delta RV$ derived from the control sample 
as an empirical 
error distribution. 

We quantified the level of radial velocity variability for each object, $x$, as the 
sum of the absolute deviation of $\Delta RV$
\begin{equation}
\label{eq:x}
x=\frac{\sum_{i=1}^{M}|\Delta\text{RV}_i|}{M}
\end{equation}
where $M$ is the number of observations of the object. We ran
a Monte Carlo simulation of $10^7$ hypothetical sets of observations 
to determine a cutoff for $x$ which is exceeded by only
$10^{-3}$ of the simulated objects. For each simulated object, 
we chose the number of
observations from the distribution of the number of observations of 
each object in the
control sample. We then randomly chose the $\Delta\text{RV}$ values 
for each simulated object from 
the 23,031 actual $\Delta\text{RV}$ values in the control sample and
calculated $x$ using Equation \ref{eq:x}. 
From the results of this simulation, we determined a cutoff value of
$x>10.4$ km s$^{-1}$ for variability significant at the $10^{-3}$ level.
Among the 1,452 objects in the experimental sample with $2\ \text{d}\le 
\Delta t\le 30\ \text{d}$, 22 exceed this cutoff and 
are therefore detected as
radial velocity variables. These stars are listed in Table \ref{data-binaries}. 

\begin{figure}[!ht]
\centering
\caption{Histograms of $\Delta RV$ and
$\Delta RV / \sigma$ for the control sample along
with the best fit Gaussian distributions. }
\label{fig:deltarv_14400}
\plottwo{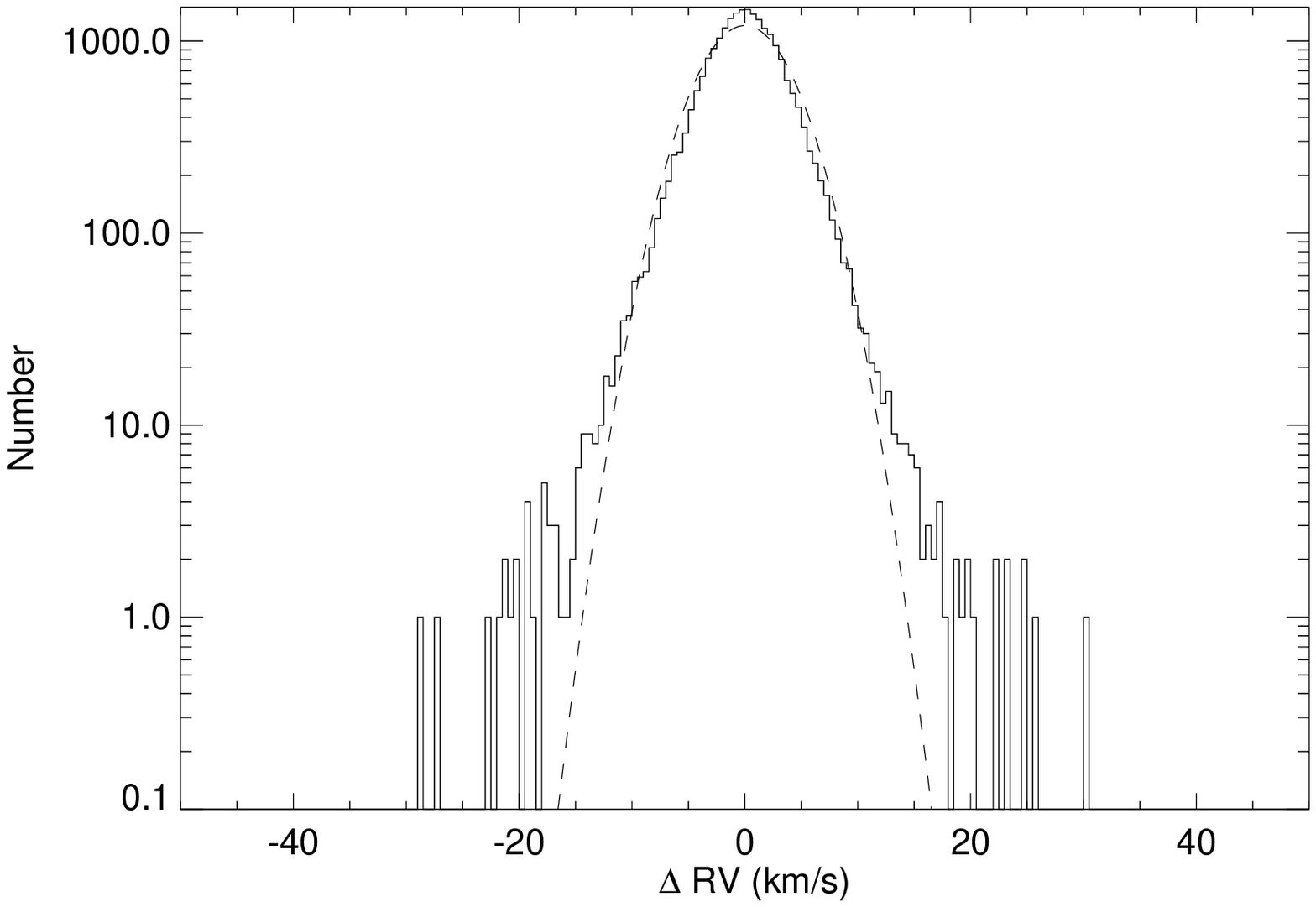}{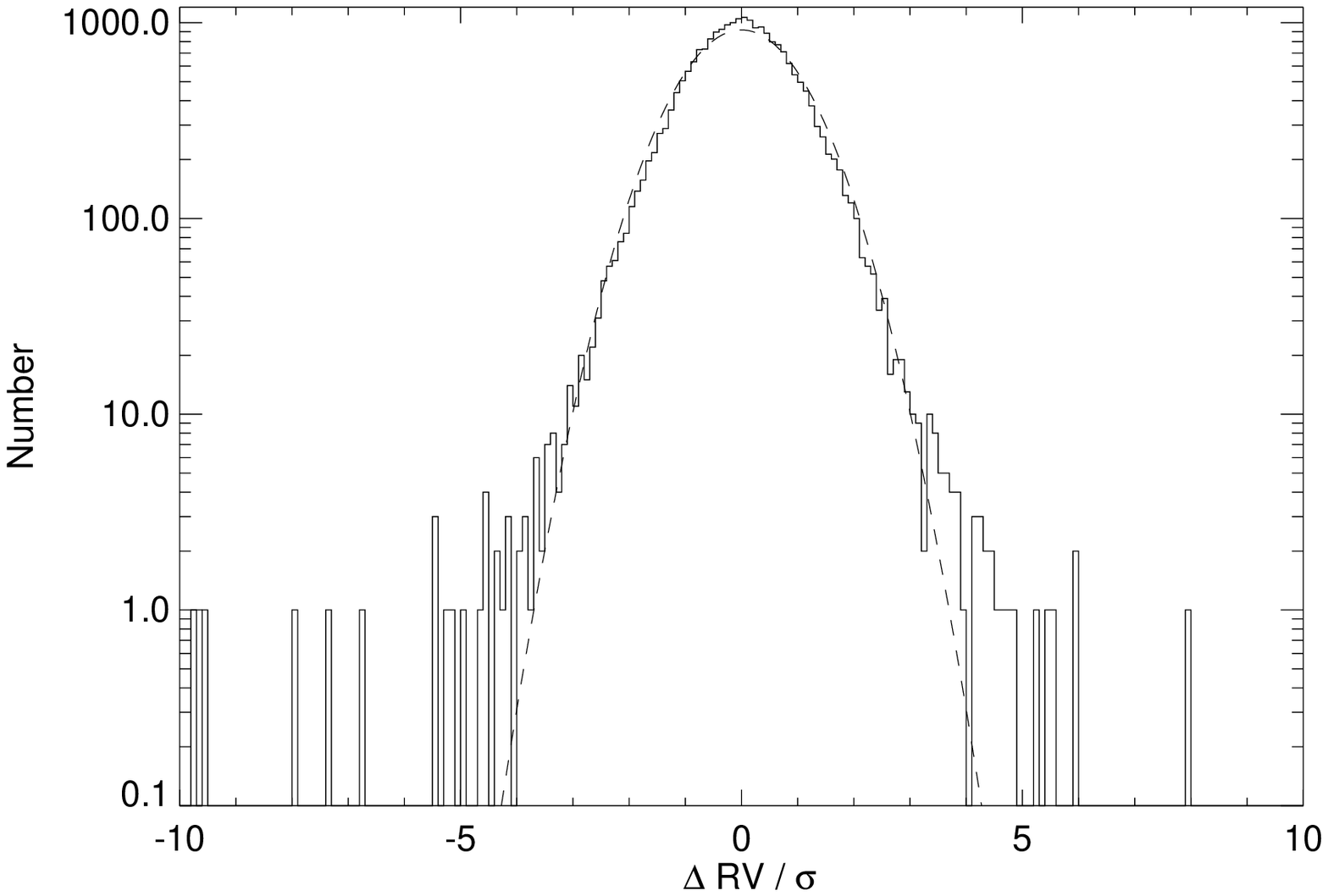}
\end{figure}

A histogram of the values of $\Delta RV$ computed for the control 
sample along with the best fit Gaussian distribution
is shown in Figure \ref{fig:deltarv_14400}. The standard deviation 
of the $\Delta RV$ values is
$3.8\ \text{km s}^{-1}$, which compares favorably with the \textit{rms} 
velocity error of
$5.5\ \text{km s}^{-1}$ at $g=18.5$ and $12\ \text{km s}^{-1}$ at $g=19.5$ 
reported by
\citet{2009ApJS..182..543A}. However,  
there is a small
set of objects with very large $\Delta RV$ values. Among the 7,059 
objects in the control sample, 
14 have at least one
observation for which $\Delta RV>20\ \text{km s}^{-1}$. 
This is far greater than
would be expected if the values of $\Delta RV$ followed a
Gaussian distribution.
We examined the spectra of these objects individually, but all 
appear to be normal
M dwarfs. However, we did note that many of the spectra contained a
small number of highly
errant pixels, likely due to cosmic rays, that were not fully 
down weighted by the SDSS pipeline.
We considered the possibility that these errant pixels were 
dramatically altering the radial velocity fit and
to test this hypothesis, we refit the spectra of the 22 detected 
binaries in the experimental sample and the 14 outliers in the
control sample with the 5 most deviant pixels removed from the fit. 
We found that the radial velocity changed
by an average of only 0.3 km s$^{-1}$, well within our radial velocity error, 
which allowed us to reject this explanation.
Template mismatch also does not appear to be at fault as the
average $\chi^2$ of these objects differs only slightly from 
the mean for the whole sample.
Based upon this, we conclude that the objects are either in fact undergoing
very short-term radial velocity variability, possibly due to a very tight 
binary system, or that
there is an error in the wavelength calibration for these spectra. 
The wavelength solutions for each spectrum taken with each SDSS fiber are determined through a combination of fits to sky emission lines and arc lamp lines. As described by \citet{2009ApJS..182..543A}, starting with DR7 these solutions were constrained to vary smoothly between fibers, resulting in overall calibration precision of $\pm 2$km s$^{-1}$ and significantly reducing the rate of errant solutions. Still, it is possible that rare problems with the wavelength solutions may be responsible for a small number of our measured RV shifts. In any case, these objects are of interest and warrant further observation.
They are listed in Table \ref{data-outliers}. We also note that a significant percentage of Sun-like stars are known to be in higher-order multiple systems (34$\%$ doubles, 9$\%$ triples; \citealt{raghavan2010}). We are only sensitive to short period systems, so even in a triple system we would only detect the reflex motion of the tight inner pair. 

The raw numbers listed above show that 22 of the 1,452 stars (1.5\%) in the 
experimental sample show statistically--significant radial velocity 
variations, compared to 14 of the 7,059 (0.20\%) of the stars in the
control sample. If the small number of large radial velocity
variations in the control sample is due to some unidentified 
instrumental or analysis problem, we would also expect 0.20\% of
the experimental sample (3 stars) to show these variations; thus 
the raw binary fraction in the experimental sample is measured at the
7$\sigma$ level (the on-average smaller number of observations 
available for the control sample is accounted for in the calculation
of $x$ in Equation \ref{eq:x}).

\subsection{The M Star Close Binary Fraction: a Bayesian Analysis}

The raw binary star fraction of 1.5\% derived above underestimates
the true binary fraction which could in principle be measured 
by radial velocities of the accuracy available from the SDSS spectra,
due to effects of inclination and the small number of observations of 
each star (see the discussion by \citealt{pourbaix05}). In this section,
we use the radial velocity measurements of the stars in the control and experimental 
samples to estimate the true close binary fraction of the 
objects in our experimental sample by quantifying the number of objects 
that exhibit statistically significant radial velocity variations. 
To do this, we take a Bayesian approach to estimate the likelihood of a
given close binary fraction given our observations and any prior information. 

Let $D$ represent the number
of radial velocity variables detected, $N_{C}$ a value for the close 
binary fraction, and $B$
knowledge of any relevant background information about the objects
and the observations of the objects, such as their magnitudes and the times of
the observations. Bayes' theorem states that
\begin{equation}
P(N_{C}|D,B) \propto P(D|N_{C},B) \cdot P(N_{C}|B)
\end{equation}
where $P(N_{C}|D,B)$, the probability of $N_{C}$ given $D$ and $B$, is
the posterior distribution; $P(D|N_{C},B)$,
the probability of $D$ given $N_{C}$ and $B$, is the likelihood 
distribution; and $P(N_{C}|B)$,
the probability of $N_{C}$ given $B$, is the prior distribution for 
any $N_{C}$ and $D$
\citep{2006DataAnalysis}.

As $D$ represents the number of objects detected as radial velocity variables, it 
follows that
\begin{equation}
\label{eq:sum}
P(D|N_{C},B)=\sum P(\{O_{i_1},O_{i_2},\ldots,O_{i_D}\}|N_{C},B)
\end{equation}
where $P(\{O_{i_1},O_{i_2},\ldots,O_{i_D}\}|N_{C},B)$ is the probability
that exactly the $D$ objects,
$i_1,i_2,\ldots,i_D$, are detected as radial velocity variables and the sum is over 
all sets of D objects. Since
whether or not
one object is detected is independent of whether or not another object
is detected,
\begin{equation}
P(\{O_{i_1},O_{i_2},\ldots,O_{i_D}\}|N_{C},B)=\prod_{j \in L} 
P(O_j|N_{C},B) \cdot
\prod_{j \not\in L} P(\bar{O}_j|N_{C},B)
\end{equation}
where $L=\{i_1,i_2,\ldots,i_D\}$, $P(O_j|N_{C},B)$ is the probability 
that the $\text{j}^{\text{th}}$ object
is detected and
$P(\bar{O}_j|N_{C},B)$ is the probability that the $\text{j}^{\text{th}}$ 
object is not detected.
Substituting this into Equation \ref{eq:sum} gives
\begin{equation}
P(D|N_{C},B)=\sum [\prod_{j \in L} P(O_j|N_{C},B) 
\cdot \prod_{j \not\in L} P(\bar{O}_j|N_{C},B)]
\end{equation}

We must now consider how to calculate $P(O_j|N_{C},B)$ and 
$P(\bar{O}_j|N_{C},B)$. Following
\citet{2005MNRAS.362L..45M}
in assuming that the only cause of radial velocity variability is stellar multiplicity,
it follows that
the probability that the $\text{j}^{\text{th}}$ object is detected as
a radial velocity variable is
$N_{C}p_{detect,j}+(1-N_{C}) \cdot 10^{-3}$ where $p_{detect,j}$ gives
the probability that we will
detect the $\text{j}^{\text{th}}$ object as a radial velocity variable if it is in
fact a binary star. Note that the term $(1-N_{C}) \cdot 10^{-3}$ arises from
the fact that $1-N_{C}$ is the probability that the object is not a binary
star while $10^{-3}$ is the probability that an object will be detected as
a radial velocity variable if it is
not a binary star by the cutoff value of the variability metric we
established to determine whether or not
an object is a radial velocity variable. Thus, $P(O_j|N_{C},B)=N_{C}p_{detect,j}+(1-N_{C})
\cdot 10^{-3}$ and
$P(\bar{O}_j|N_{C},B)=1-P(O_j|N_{C},B)=1-[N_{C}p_{detect,j}+(1-N_{C})
\cdot 10^{-3}]$. We must now determine
$p_{detect,j}$ for each of the objects in the experimental sample, which
can be done with a Monte Carlo simulation.

The radial velocity measurements described in section \ref{sec:id} and models for
the distributions of system orbital parameters formed the basis of a Monte Carlo 
simulation designed to determine $p_{detect,j}$, which is
required to generate the posterior distribution. This Monte Carlo
simulation consisted of $10^5$ virtual binary stars for each of the 1,452
objects in the experimental sample, which enabled us to estimate
the efficiency with which we detect binaries 
given a wide range of different binary systems. For this simulation, 
we drew binary parameters from distributions based on previous results 
in the literature. 

{\bf Semi-major axis, $a$}: In order to have a reasonable chance to detect a
binary in the experimental sample, there must be significant RV variations on the timescale of $\Delta t$, typically less than five days \citep{pourbaix05}. Since the stars we are looking at are
M stars, $m_1\sim m_{2}<0.5\ M_{\odot}$, and we are able to detect large RV variations, say $\Delta RV > 20 $ km s$^{-1}$,
we are primarily sensitive to very close systems with $a<0.2\ \text{AU}$.
Based upon our Monte Carlo simulations, we estimate that our average detection
efficiency is 69\% at $a=0.1\ \text{AU}$ and 16\% at $a=0.4\ \text{AU}$.
The distribution of the semi-major axes for systems with such
small separations is not well known. We consider two different 
distributions of $a$. The first is a
uniform distribution from 0.01 to 0.4 AU, while the second 
is a linear
distribution such that $P(a) \propto a$ that also runs from 0.01 to 0.4 AU.
Note that we might expect the uniform distribution to overestimate
the value of $p_{detect,j}$ as the correct
distribution should have a higher probability of larger separations
and a corresponding
lower probability of smaller separations \citep{2007ApJ...668..492A}. 
This overestimate of
$p_{detect,j}$ in turn implies
that the best fit-binary fraction $N_{C}$ will be
underestimated\label{n_underestimate}.

{\bf Mass ratio, $q$}: We follow \citet{2007ApJ...668..492A}
in using a power law distribution with
a minimum
of $q=0.02$. Thus, if we let $\gamma$ represent the power law index,
the probability distribution
function of $q$ is
\begin{equation}
P(q)=\frac{q^{\gamma}}{\int_{0.02}^1 q^{\gamma}\,\mathrm{d}q}
\end{equation}
for $.02<q<1$ and $P(q)=0$ for $0<q<0.02$. We test distributions 
with three different
values for $\gamma$. We test $\gamma=1.8$, as found by
\citet{2007ApJ...668..492A}, and
$\gamma=1.2$ and 2.2, the extreme values on the $1\sigma$ confidence 
interval given by
\citet{2007ApJ...668..492A}.

{\bf Primary mass, $m_1$}: Unfortunately, there is not a good method 
for determining the mass of the
primary from the SDSS spectra as there are no well-calibrated 
mass-color or mass-luminosity
relationships using the SDSS filters. Additionally, not knowing the
metallicity or age of the star increases the uncertainty in determining
its mass. However, this is not a major issue because, while the 
optical spectra and brightness of cool M dwarfs cover a large range,
the corresponding mass range is modest, and
we are able to obtain a rudimentary estimate of the primary mass
using relations from the literature. Using the color transformations
provided by \citet{2006PASP..118.1679D},
from the apparent magnitude of the star in the $r$ and $i$ bands we
can estimate the $i-J$ color. Also, using the color-magnitude 
relationships provided by
\citet{2005PASP..117..706W}, we can determine the absolute magnitude, $M_i$
from the $i-z$ color of the objects. Since $i-J=M_i-M_J$, we 
can estimate $M_J$. Finally,
the mass-luminosity relations of \citet{2000A&A...364..217D} 
allow for the estimation of the star's
mass from $M_J$. While the individual values of the mass obtained
by these means have large systematic uncertainties, this is mitigated 
by selecting primary mass from a uniform distribution from 0.75
to 1.25 times the photometrically estimated mass in our Monte Carlo simulations. 

{\bf Eccentricity, $e$}: It is known that binaries with very short
periods (P$<\sim$10d) are
highly likely to undergo tidal circularization 
\citep{1991A&A...248..485D,2005ApJ...620..970M,raghavan2010}
and we can reasonably assume circular orbits, although the 
largest--separation pairs may have
$e \ne 0$.

{\bf Orbital phase, $f$}:
The orbital phase at the time of the first observation is chosen 
from a uniform
distribution from 0 to $2\pi$.

{\bf Inclination, $i$}:
The inclination is chosen from a uniform distribution from 0 
to $\pi$ radians.

{\bf Longitude of periastron, $\omega$}:
The longitude of periastron is chosen from a uniform distribution 
from 0 to $2\pi$ radians.

With our Monte Carlo simulation we generate hypothetical radial velocity curves for 
binary systems given the actual
times of the observations of each object and Keplerian orbits given 
randomly selected system parameters from the above distributions. 
We estimate errors on the individual data points in the simulated radial velocity 
curves by randomly selecting values from the 
$\Delta RV$ distribution of our control sample. We can then compute 
$x$ using Equation \ref{eq:x}. If the value of $x$ is greater than 
the cutoff value
determined in section \ref{sec:id}, $10.4$ km s$^{-1}$, this system is 
tagged as detected. By running $10^5$ trials for each of the 
1,452 objects in our experimental sample
and determining the fraction of trials in which we detect 
the simulated object, we estimate 
$p_{detect,j}$ as a function of the binary system parameters. 

The prior distribution, $P(N_{C}|B)$, allows us to incorporate any 
relevant outside information into our analysis. 
We follow \citet{2007ApJ...668..492A} in choosing a prior that assumes
no outside knowledge and
therefore is not biased towards any particular values of $N_{C}$. Thus,
as $N_{C}$ is a scale parameter,
the proper prior to use is the Jeffreys' prior 
\citep{2006DataAnalysis}, 
\begin{equation}
P(N_{C}|B) \propto \frac{1}{N_{C}}\text{.}
\end{equation}
which is equivalent to uniform in the log of $N_{C}$.

Given the likelihood, $P(D|N_{C},B)$, and prior distribution, 
$P(N_{C}|B)$, we can calculate the posterior distribution: 
\begin{IEEEeqnarray}{rCl}
P(N_{C}|D,B) & \propto & P(D|N_{C},B) \cdot P(N_{C}|B) \\
& \propto & \sum [\prod_{j \in L} P(O_j|N_{C},B) \cdot 
\prod_{j \not\in L} P(\bar{O}_j|N_{C},B)]
\cdot \frac{1}{N_{C}} \\
\label{eq:posterior}
& \propto & \sum [\prod_{j \in L} [N_{C}p_{detect,j}+(1-N_{C})10^{-3}] \cdot
\prod_{j \not\in L} [1-(N_{C}p_{detect,j}+(1-N_{C})10^{-3})]]
\cdot \frac{1}{N_{C}}
\end{IEEEeqnarray}
We can then determine the constant of proportionality using the 
normalization condition
\begin{equation}
\int_0^1 P(N_{C}|D,B)\,\mathrm{d}N_{C}=1.
\end{equation}
From this posterior distribution, we are able to
determine the best fit value of the close binary fraction and 
a confidence interval on this value.

\begin{figure}[!ht]
\centering
\caption{Posterior distributions for
a uniform (solid line) and linear (dashed line) distribution of $a$, respectively.}
\label{fig:posterior}
\includegraphics[scale=.8]{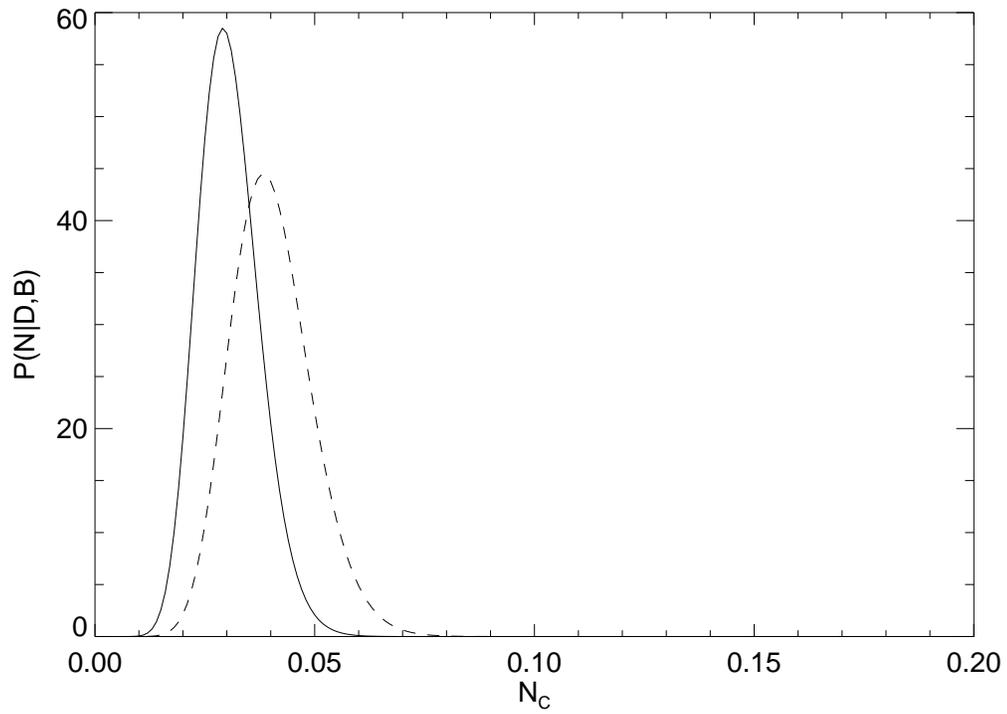}
\end{figure}

Table \ref{tab:best_n}
gives the average value of $p_{detect,j}$ over all the objects 
and the best fit value
of $N_{C}$ and the $1\sigma$ confidence interval on $N_{C}$ 
for each combination of $a$ and
$q$ distributions. From Table \ref{tab:best_n}, it is clear that 
changing the power law index
$\gamma$ has negligible impact on $N_{C}$. However, using a linear 
distribution for $a$ as opposed to
a uniform distribution increases $N_{C}$ as expected, due to 
the decrease of $p_{detect,j}$
that results from using a distribution with a higher probability
of large separations.
The increase in $N_{C}$ 
that results from using
a linear distribution instead of a uniform distribution is $0.9\%$.
Figure \ref{fig:posterior} shows the posterior distributions for
the close binary fraction for both a uniform and linear distribution
of $a$.
Assuming the uniform distribution in $a$, the close binary fraction
of the objects in our sample is $2.9^{+0.6}_{-0.8}\%$, while for a linear distribution in $a$ we find $3.8^{+0.9}_{-0.9}\%$. We note that since our sample is magnitude limited, Branch bias \citep{branch1976} may lead us to overestimate the rate of occurrence of multiple systems. For a population of binary systems with two identical stars, this can bias the measured binary fraction by up to $35\%$. We have not taken this into account in our analysis. 

\section{Results and Discussion}
\label{sec:resultsdiscuss}

Previous work on the statistical properties of multiple star systems, 
such as that by \citet{1991A&A...248..485D} and \citet{raghavan2010}, 
has largely excluded M stars. \citet{1992ApJ...396..178F} included 
spectroscopic binaries in their analysis of M star multiplicity and 
estimated that $\sim 1.8\%$ of
M dwarfs are binaries with $0.04\ \text{AU}<a<0.4\ \text{AU}$. This 
result, given that it is based on a very small sample of targets, 
is fully consistent with our estimate of the binary fraction across 
a similar range of separations.   
The rate of occurrence of multiple systems, and the dependence 
of that rate on primary mass, may be an important observational 
clue to a unified
theory of star formation. It has been shown by \citet{lada2006} 
and \citet{raghavan2010} that the overall multiple fraction is a 
strong function of stellar mass, decreasing from near $80\%$ at 
10$M_{\sun}$ to $20\%$ for brown dwarfs with masses $<0.1$M$_{\sun}$.
We gathered values from the literature for the binary and close--binary
fractions as a function of primary mass. These are listed in Table \ref{tab:fraction}.

\begin{figure}[!ht]
\centering
\caption{Values of the close binary fraction and total binary fraction given in
Section \ref{sec:resultsdiscuss} and shown in tabular format in Table \ref{tab:fraction} along with best
fit line for each fraction and total binary fraction.}
\label{fig:binfrac}
\includegraphics[scale=.8]{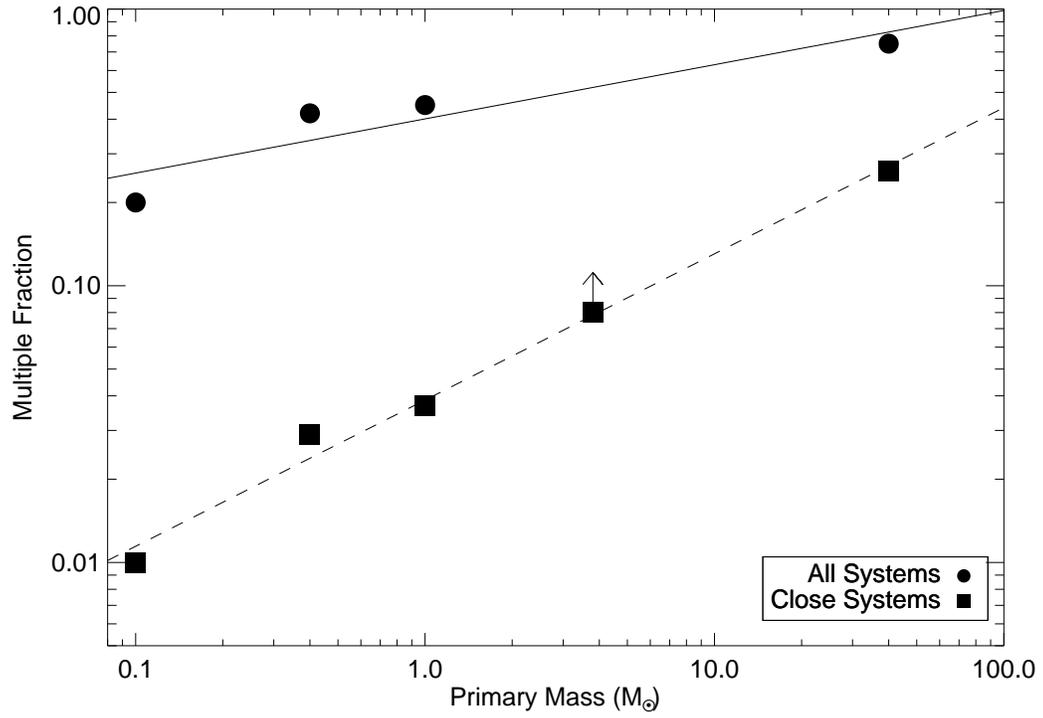}
\end{figure}

For the overall multiple fraction, $N_{T}$, we take the results of
\citet{2007ApJ...668..492A} for the brown dwarfs and lowest mass stars,
\citet{1992ApJ...396..178F} for M stars, \citet{raghavan2010} for G stars,
and \citet{mason2009} and \citet{2010arXiv1009.4197S} for O stars,
for which we assume an average 
primary mass of $40$M$_{\sun}$ based on the studies of O star eclipsing
binaries by \citet{weiden2010}. For the close binary fraction,
$N_{C}$, of the lowest mass stars and brown dwarfs, we take the results
from \citet{2010ApJ...723..684B} for the semi-major axis
range $a<1$ AU and scale down by a factor of 0.4 to estimate the binary
fraction in the range $a<0.4$ AU. For M stars we take the results 
presented here. For G stars we integrate the overall semi-major axis 
distribution from \citet{raghavan2010}, inferred from the period 
distribution by assuming binary systems with a total mass of 
1.5M$_{\sun}$, for $a<0.4$AU. For B stars we take the lower limit on the rate of occurrence of spectroscopic binaries with periods less the 10d derived by \citet{mazeh2006} by applying the most extreme correction for the Branch bias to the result of \citet{wolff1978} and assume a stellar mass of 3.8M$_{\sun}$. Finally, for O stars we use the empirical Cumulative Distribution
Function (CDF) of observed O star spectroscopic binary periods to
estimate $N_{C}$ for $a<0.4$AU assuming the binaries have a total mass
of 70M$_{\sun}$, and therefore periods less than $10$d. The relation
between $N_{T}$ and $N_{C}$ and primary mass is shown in
Figure \ref{fig:binfrac}. 

These results show that the binary fraction is 
lower for lower mass stars across a wide range of orbital separations. However, the ratio $N_{C}$/$N_{T}$ appears to increase with increasing primary mass, indicating that either the peak of the semi-major axis distribution moves to smaller values of $a$ for larger stellar masses or that the overall shape of the separation distribution evolves significantly with primary mass. Based on well-studied samples of M/L and G stars, there is evidence that in fact the lowest mass binaries tend to have smaller separations \citep{1991A&A...248..485D,2007ApJ...668..492A, raghavan2010}. We caution that for O and B stars the multiplicity studies are not as observationally complete as those focusing on L and G stars, though with such high multiple fractions for these massive stars already reported in the literature it is unlikely that a significant population of wide separation systems remains undetected. Even if, as suggested by \citet{kroupa2011}, there is a universal Initial Period Function (IPF) for all binary stars, the processes of dynamical destruction by stars external to bound stellar systems
and dynamical decay, i.e. ejection by stars internal to the stellar system, must play an important role in the evolution of binary properties (\citealt{1975MNRAS.173..729H}, \citealt{1975AJ.....80..809H}) and a wide range of binary properties for field stars may be expected.

\section{Conclusions}

Using a large set of spectroscopic observations of cool dwarf M stars from
the SDSS that provide sparsely-sampled time series (typically 3-4 
observations) we isolate 22 close spectroscopic binary candidates by radial
velocity variability. Comparison with the radial velocity distributions
of a control sample also from SDSS shows that we have detected the raw binary
fraction at a 7$\sigma$ level. The cadence of the observations, and the 
SDSS radial velocity accuracy, make us sensitive to close binaries, with
separations less than about 0.4 AU. The detectability of spectroscopic binaries in the SDSS observations is
evaluated using a large Monte Carlo simulation of the observational properties
of the underlying population. Taking into account this detection efficiency and the total number of objects in our sample, the close binary fraction ($a<0.4\ \text{AU}$) of dwarf M stars
is determined to be $2.9^{+0.6}_{-0.8}\%$ assuming a uniform prior on the distribution of orbital separation $a$. Our estimated close binary fraction of $3-4\%$, depending on the prior for $a$, is broadly consistent with earlier results on the statistical properties of low-mass multiple systems. 

By comparing our measurement of the close binary fraction of M dwarfs to previous results in the literature, we have shown that the
close binary fraction, like the overall binary fraction,
is an increasing function of primary mass. This result has implications for
the formation of, and the early dynamical evolution of, systems of low
mass stars and indicates that the overall distribution of binary separations may be a strong function of stellar mass. In the future, the methods described in this work can be used to investigate the
binary fraction in the whole sample of stars observed spectroscopically
by SDSS -- 
over 460,000 stars, mostly of types K, G and F,
and members of the thick disk and halo.

\section{Acknowledgments}
The authors would like to thank an anonymous referee for thoughtful comments that helped to improve this manuscript. 
We also thank Jim Gunn, Craig Loomis, Steve Bickerton, Fergal Mullally and Robert
Lupton for their extensive work on the SDSS spectroscopic pipeline and
in particular for making available the calibrated individual 15-minute 
spectra. We also thank David Latham for helpful converstations that contributed to this work. CHB thanks the National Science Foundation for support via 
NSF postdoctoral fellowship AST-0901918, and GRK thanks the NSF and NASA
for support via grants NNX07AH68G and AST-0706938. We thank Princeton's
astrophysics department, in particular David Spergel, for strong support of 
the undergraduate research program, which supported this work.

Funding for SDSS and for SDSS-II was provided by the Alfred P.
Sloan Foundation, the Participating Institutions, the National
Science Foundation, the U.S. Department of Energy, the National
Aeronautics and Space Administration, the Japanese Monbukagakusho,
the Max Planck Society, and the Higher Education Funding Council
for England.  The SDSS is managed by the Astrophysical Research
Consortium for the Participating Institutions.

\newpage

\begin{deluxetable}{lrrrr}
\tabletypesize{\scriptsize}
\tablecaption{Object name,
i magnitude, value of $x$ calculated using Equation \ref{eq:x}, the range in
$RV$ values ($RV_{range}$=$RV_{max}-RV_{min}$), and $\Delta t$ for each object
in the sample with $2\ \text{d}\le \Delta t\le 30\ \text{d}$ that we detected as a binary.\label{data-binaries}}
\tablewidth{0pt}
\tablehead{
\colhead{Object Name} & \colhead{i} & \colhead{$x$} & \colhead{$RV_{range}$} & \colhead{$\Delta t$}\\
\colhead{} & \colhead{} & \colhead{(km/s)} & \colhead{(km/s)} & \colhead{(d)}
}
\startdata
SDSSJ002228.76$-$010806.9	& 16.59&      12&      38&      19.002\\
SDSSJ004747.84$-$004550.5	& 17.02&      17&      38&      17.938\\
SDSSJ010122.67$-$005311.6	& 17.07&      14&      35&       2.019\\
SDSSJ011038.87$-$011409.5	& 17.31&      12&      32&       6.013\\ 
SDSSJ011054.12+005227.4	& 16.54&      14&      37&       6.013\\
SDSSJ074428.29+191554.7	& 17.47&      70&     251&       5.035\\
SDSSJ074646.12+282629.9	& 17.55&      12&      44&       7.056\\
SDSSJ083027.95+454736.5	& 17.33&      13&      35&       4.184\\
SDSSJ083723.40+141211.1 & 16.95&      11&      26&      14.936\\
SDSSJ084841.17+232051.7	& 16.34&      92&     202&       8.090\\
SDSSJ105030.21+421451.4	& 16.44&      24&      90&       3.004\\
SDSSJ105715.76+430945.9	& 16.33&      24&      97&       3.015\\
SDSSJ114030.06+154231.5	& 16.18&      47&      98&      17.963\\
SDSSJ114050.85+532304.0	& 16.65&      13&      35&      24.966\\ 
SDSSJ115124.34+371953.8	& 16.37&      14&      40&       2.135\\
SDSSJ121944.13+260759.8	& 16.13&      12&      36&       2.865\\
SDSSJ163215.69+005918.6	& 16.66&      24&      98&      24.994\\
SDSSJ163401.41+005010.0	& 16.87&      14&      53&      24.994\\
SDSSJ204845.85+004001.3	& 17.47&      47&     182&      19.957\\
SDSSJ212546.00$-$060858.6	& 16.12&      16&      43&      22.955\\
SDSSJ220848.44+000409.9	& 17.12&      23&      59&      20.869\\
SDSSJ225450.30$-$101003.2	& 17.17&      17&      40&      21.975\\

\enddata
\end{deluxetable}

\begin{deluxetable}{lrrrr}
\tabletypesize{\scriptsize}
\tablecaption{Object name,
i magnitude, value of $x$ calculated using Equation \ref{eq:x}, the range in
$RV$ values ($RV_{range}$=$RV_{max}-RV_{min}$), and $\Delta t$ for each object
in the sample with $0\ \text{hr}\le \Delta t\le 4\ \text{hr}$ with at least one observation
for which $\Delta RV>20\ \text{km/s}$.\label{data-outliers}}
\tablewidth{0pt}
\tablehead{
\colhead{Object Name} & \colhead{i} & \colhead{$x$} & \colhead{$RV_{range}$} & \colhead{$\Delta t$}\\
\colhead{} & \colhead{} & \colhead{(km/s)} & \colhead{(km/s)} & \colhead{(d)}
}
\startdata

SDSSJ003144.47+003033.6	& 17.49&      14&      40&     0.758\\
SDSSJ021838.39+004946.8	& 16.98&      16&      41&     0.685\\
SDSSJ072642.62+414243.2	& 16.64&      14&      44&     1.188\\
SDSSJ075243.70+254928.3	& 17.39&      15&      41&     0.723\\
SDSSJ082833.58+341531.6	& 16.79&      15&      35&     0.554\\
SDSSJ085921.75+371147.9	& 18.19&      15&      34&     1.087\\
SDSSJ092345.54+222432.4	& 16.56&      48&     133&     1.242\\
SDSSJ111647.81+294602.7	& 16.24&      14&      36&     0.585\\
SDSSJ125508.85+320849.9	& 17.04&      10&      33&     1.053\\
SDSSJ143524.64+232249.5	& 16.97&      17&      48&     0.564\\
SDSSJ154848.35+362803.7	& 17.04&      17&      42&     0.438\\
SDSSJ163020.19+305254.5	& 16.62&      20&      55&     1.343\\
SDSSJ163355.96+293725.0	& 16.98&      13&      34&     1.343\\
SDSSJ163544.45+243032.3	& 16.78&      13&      35&     1.873\\

\enddata
\end{deluxetable}

\begin{deluxetable}{cccc}
\tablecaption{$\langle p_{detect,j}\rangle$, the best fit value of $N_{C}$ and $1\sigma$
(68.3\%) confidence interval on $N_{C}$ for each combination of $a$ and $q$ distributions.\label{tab:best_n}}
\tablewidth{0pt}
\tablehead{
\colhead{P($a$)} & \colhead{P($q$)} & \colhead{$\langle p_{detect,j}\rangle$} & \colhead{$N_{C}$}\\
}
\startdata
Uniform & $\gamma=1.2$ & 0.45 & $3.0^{+0.5}_{-1.0}\%$ \\
Uniform & $\gamma=1.8$ & 0.46 & $2.9^{+0.6}_{-0.8}\%$ \\
Uniform & $\gamma=2.2$ & 0.47 & $2.9^{+0.5}_{-0.9}\%$ \\
Linear & $\gamma=1.2$ & 0.34 & $4.0^{+0.8}_{-1.1}\%$ \\
Linear & $\gamma=1.8$ & 0.35 & $3.8^{+0.9}_{-0.9}\%$ \\
Linear & $\gamma=2.2$ & 0.36 & $3.8^{+0.8}_{-1.0}\%$ \\
\enddata
\end{deluxetable}

\begin{deluxetable}{rrr}
\tablecaption{Summary of current results for the total binary fraction and close binary
fraction for stars of various primary masses. See Section \ref{sec:resultsdiscuss} for information
on the sources of these values. \label{tab:fraction}}
\tablewidth{0pt}
\tablehead{
\colhead{Primary Mass} & \colhead{$N_{T}$} & \colhead{$N_{C}$}\\
\colhead{(M$_{\sun}$)} & \colhead{($\%$)} & \colhead{($\%$)}
}
\startdata
0.1 & 20  & 1.0 \\
0.4 & 42 & 2.9 \\
1.0 & 45 & 3.7  \\
3.8 & $-$ & $>$8.0\\
40.0 & 75 & 26.0 \\
\enddata
\end{deluxetable}

\end{document}